\documentclass[12pt]{article}
\usepackage{graphicx}
\usepackage[round]{natbib}
\usepackage{epsfig}
\title{
Using the QBO to predict the number of hurricanes hitting the U.S.
 }

\author{
Katie Coughlin (RMS)\footnote{\emph{Correspondence email}: \texttt{katie.coughlin@rms.com}}\\
}

\begin{document}
\maketitle

\begin{abstract}
\noindent
A simple study of the relationship between the QBO and the number of hurricanes in the Atlantic, both in the Basin and hitting the U.S. coastline, demonstrates that the QBO is not a particularly useful index to help predict hurricane numbers on five-year time scales.  It is shown that there is very little difference between the number of hurricanes following easterly winds in the equatorial stratosphere and the number that follow westerly winds.  Given this it is reasonable one would make better predictions just using the mean number of hurricanes in lieu of using the QBO and this is also simply demonstrated here.
\end{abstract}

\section{Introduction}
Hurricanes have a severe impact on coastal communities in the U.S. and these communities are justifiably concerned and interested in the prediction of the number of forthcoming hurricanes. Our particular interest is in making quantitative predictions of the number of hurricanes that might occur in the next five years and this paper investigates the usefulness of the Quasi-Biennial Oscillation (QBO)when making such a prediction.  \\
\\
The QBO is a stratospheric phenomena.  Winds in the tropical stratosphere (from about 14 km and up to about 80 km, which is in the mesosphere) alternate between easterlies and westerlies with an average period of about 28 months.  Alternating winds descend from above at an average rate of 1km/month but as they near the tropopause the regularity of this oscillation weakens and there is little influence below the tropopause.  In general, these winds blow in a swath with width $15^{\circ}$ on both sides of the equator.  And, although these winds have a very large effect in the upper atmosphere, the mechanisms by which they might affect the troposphere are not very well understood.\\
\\
Gray et al. [1992] and Shapiro et al. [1987], however, find that the frequency of tropical cyclones in the Atlantic basin are sensitive to the phase of the QBO.  Gray [1984] claims that this may have to do with static stability and a dynamical interaction near the tropopause.\\
\\
Here we see if we can use this sensitivity to gain skillful predictions of the average number of hurricanes in the next 5 years.  First we test to find which months and which levels of the QBO are most sensitive to hurricane numbers (by performing lag correlations).  Note that this may not be the best way to determine sensitivity but since the wind speeds do not vary too much in height or month, this method gives us a reasonable starting point.  The QBO is then divided into easterly and westerly years.  Very few of the years are "neutral" because the QBO tends to act more like a square wave in time since the transitions from east to west and back again occur much quicker than the period of the oscillation.  The average number of hurricanes in the years following these easterly or westerly QBOs are then grouped and boxplots are used to distinguish whether or not the groups are significantly different.  As a further test, these groups are also used as an estimate for predicting the number of hurricanes following future easterly or westerly QBOs.  Using an out-of-sample hindcasting experiment, the past numbers of hurricanes are predicted according to the mean number of storms following past easterly or westerly QBOs and these predictions are compared to the actual numbers of storms which occurred to calculate an error.  The out-of-sample root mean squared error (RMSE) is calculated and is compared to the out-of-sample RMSE for a climatological prediction. \\

\section{Data}
The hurricane record that we use comes from HURDAT [Jarvinen et al.,1984].  For landfalling storms we only count the number of first landfalls on the U.S. coast.  Reliable intensity measurements date back to when aircraft were first used for reconnaissance, 1944 [Neumann et al., 1993] and for U.S. landfalling hurricanes, accurate measures of central pressure at landfall are reliable back to 1899 [Jarrell et al., 1992].  However, we use Barbara Naujokat's radiosonde wind speeds [Naujokat et al., 1986] for the stratospheric QBO and those measurements are only available since 1953.  It is this date then, that constrains the following analyses.

\section{Results}
We first take a look at the relationship between the QBO and hurricane numbers.  Figures 1 and 2 show the correlation between the numbers of hurricanes each year and the QBO wind velocity at various heights in the stratosphere (70-10 hPa) and at each month before the hurricane season.  The January to August QBO values correspond to the wind velocity of the QBO occurring during the same year as the hurricane season and the September to December values correspond to the QBO one year before the hurricane season that it is being correlated with.  Figure 1 shows correlation values between the number of storms (cat 1-5 in the top row and cat 3-5 in the bottom row) in the Atlantic basin and the QBO winds at various heights in the stratosphere.  Figure 2 shows similar correlations for the number of hurricanes which hit the U.S. coastline.  Kendall's tau statistic is used to estimate a rank-based measure of association that is robust and suitable for data that does not necessarily come from a bivariate normal distribution.  Red colors denote negative correlations and white values denote positive correlations.  The largest correlations are described in the sub-header of each plot.  The diagonal structure in the correlation, descending from left to right, which can be picked out by eye is due to the fact that the phase of the QBO descends at about 1km/month.  However, none of the correlations shown are significantly different from zero (at a 95\% confidence level) according to the test for association between paired samples as described in Hollander and Wolfe [1973].\\
\\
\begin{figure}[!hb]
  \begin{center}
    \scalebox{0.7}{\includegraphics{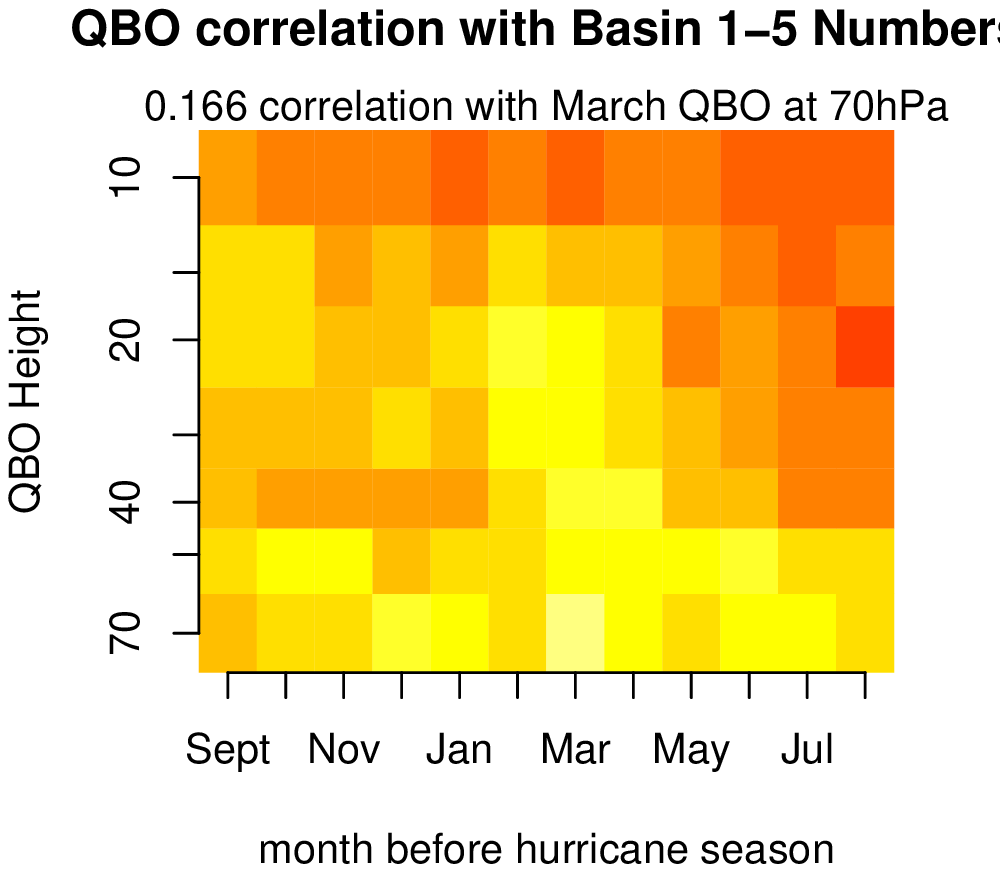}}
    \scalebox{0.6}{\includegraphics{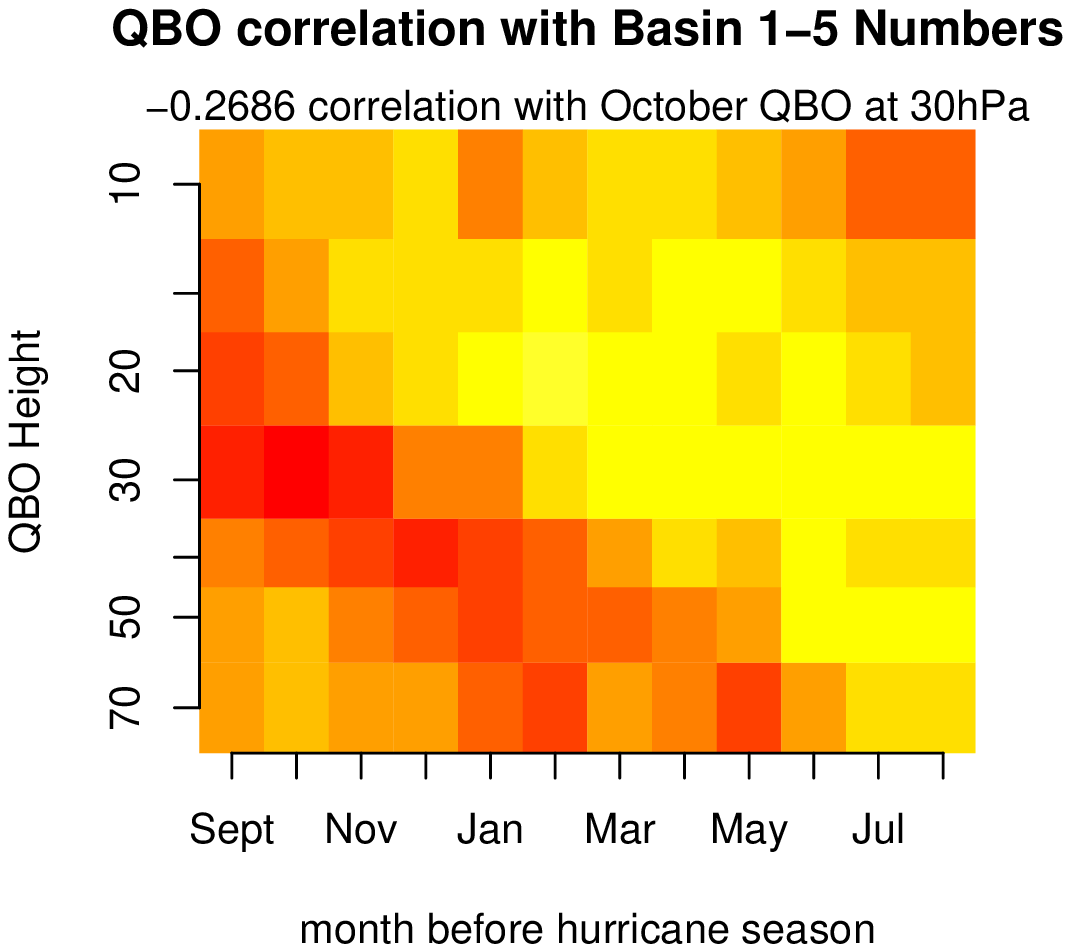}}
    \scalebox{0.7}{\includegraphics{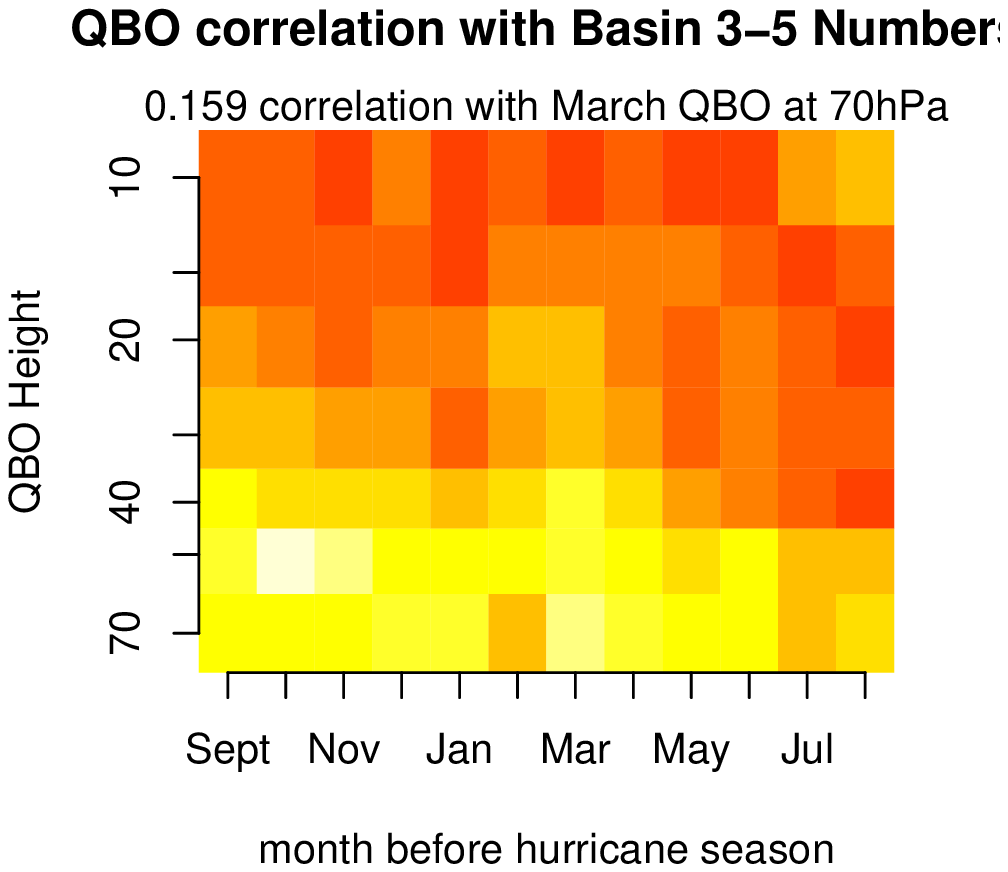}}
    \scalebox{0.6}{\includegraphics{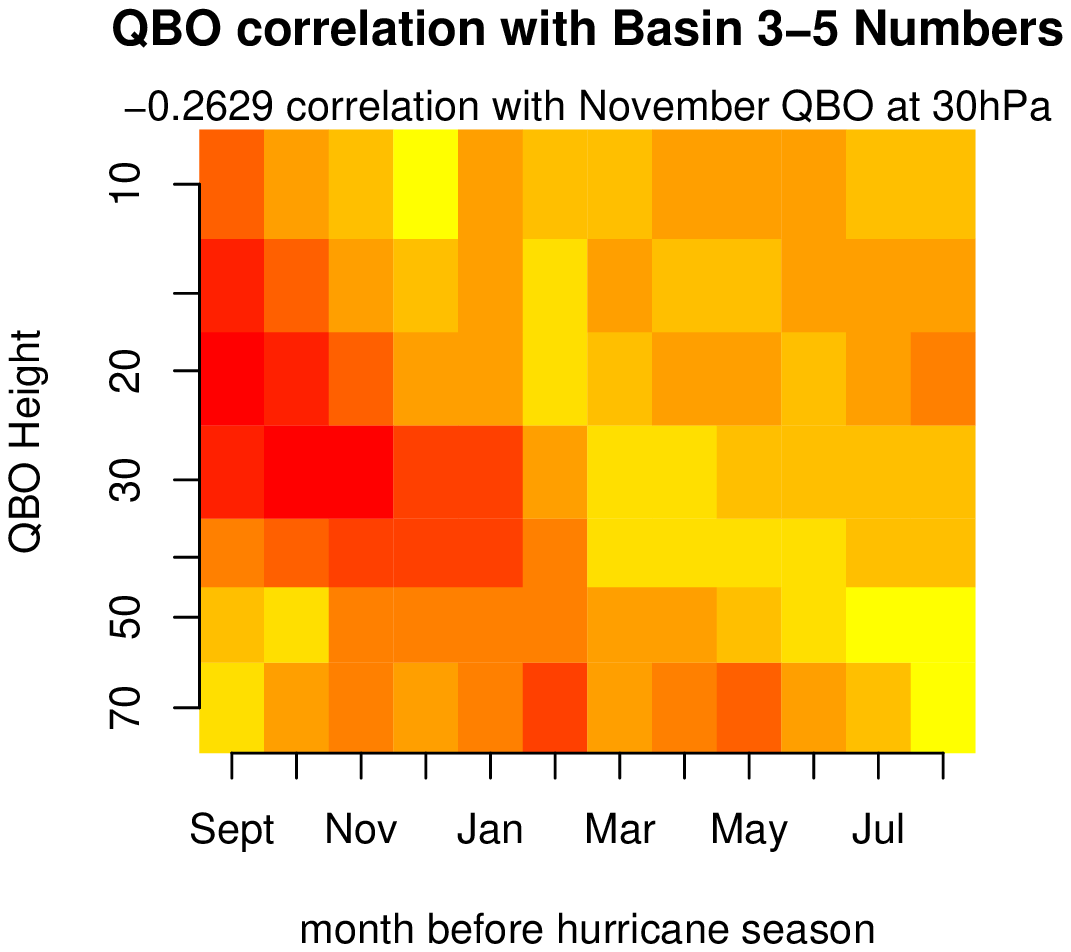}}
  \end{center}

    \caption{\textbf{Correlations between the QBO and the number of hurricanes in the Basin} (cat 1-5 is shown in the top two panels and cat 3-5 is shown in the bottom two).  Red denotes negative correlations and white positive correlations.  Correlations are calculated from 1953 using all of the QBO data (shown on the left) and using data just from the second half of the time series, 1979 (shown on the right).  The largest correlations for each figure are mentioned in the subheader of each figure.
}
     \label{f01}
\end{figure}

\begin{figure}[!hb]
  \begin{center}
    \scalebox{0.7}{\includegraphics{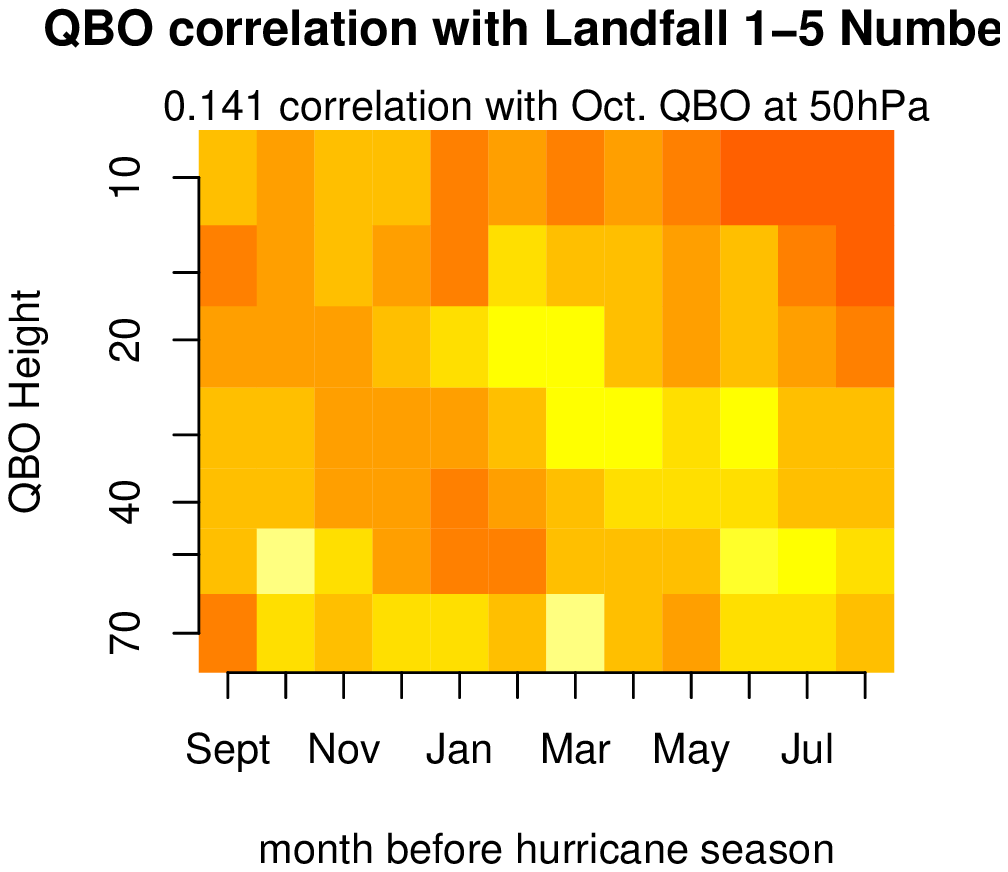}}
    \scalebox{0.6}{\includegraphics{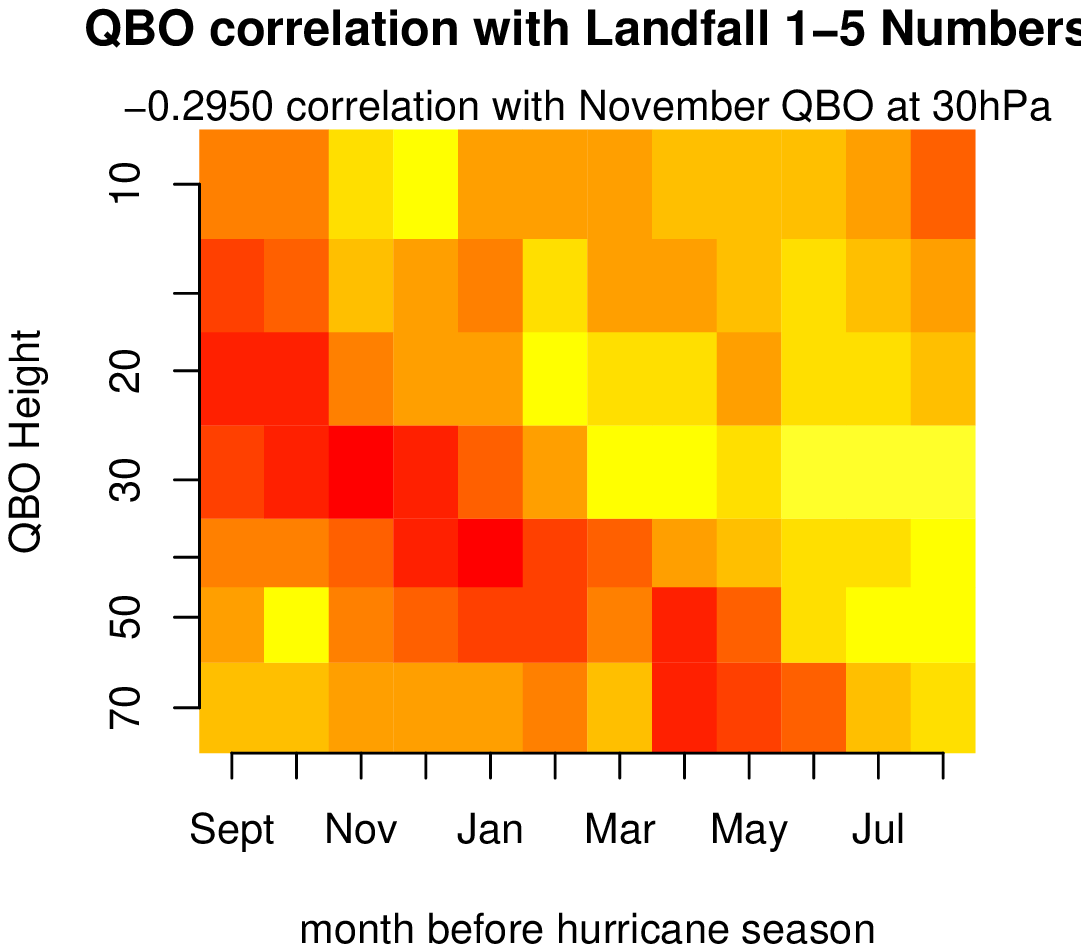}}
    \scalebox{0.7}{\includegraphics{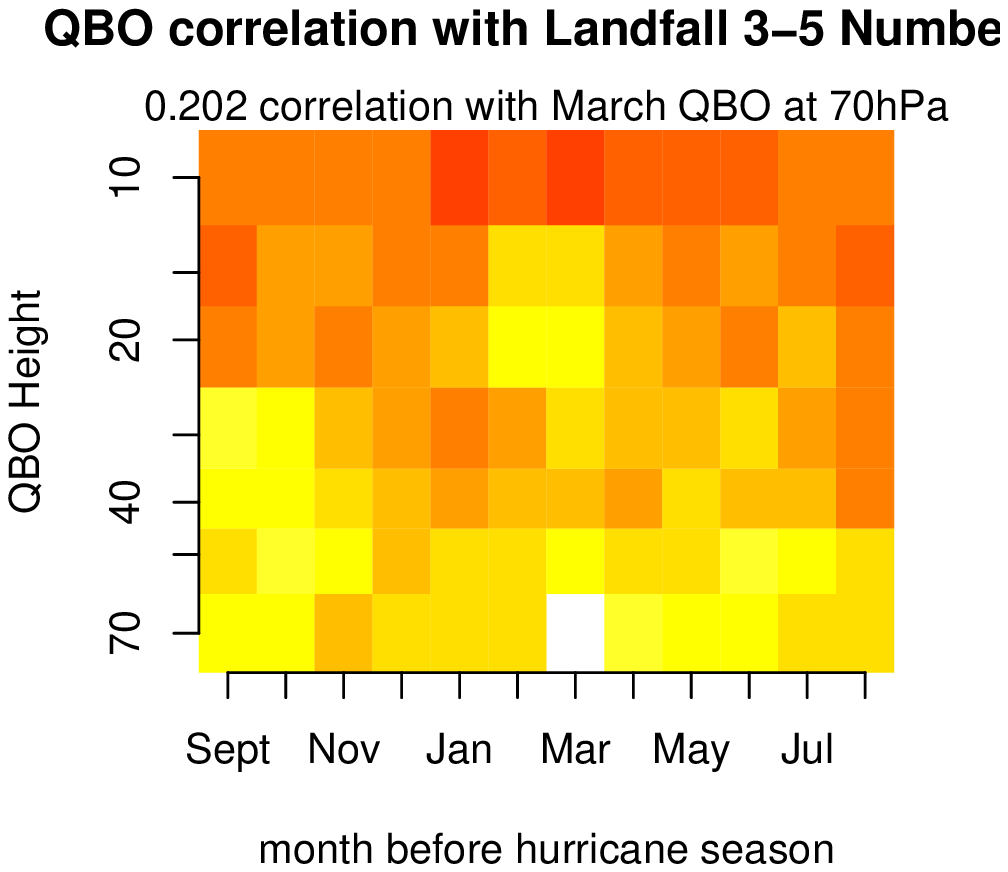}}
    \scalebox{0.7}{\includegraphics{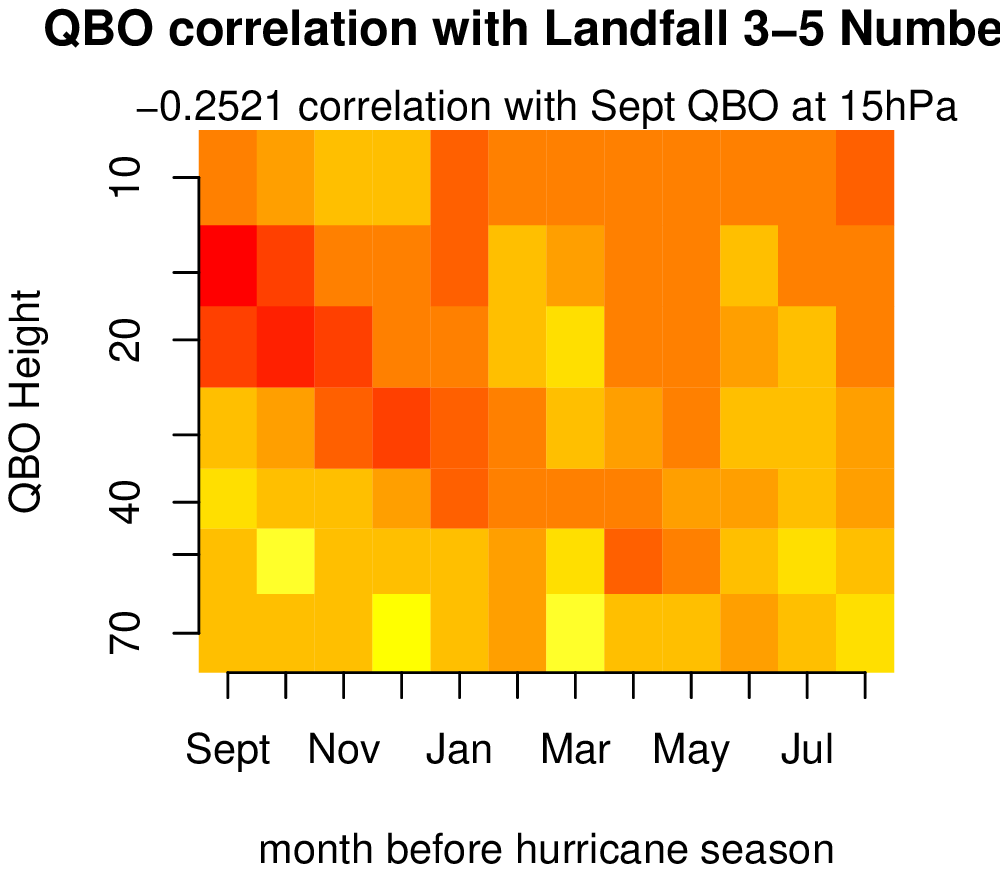}}
  \end{center}
    \caption{\textbf{Correlations between the QBO and the number of hurricanes which hit the U.S. coast} (cat 1-5 is shown in the top two panels and cat 3-5 is shown in the bottom two).  Red denotes negative correlations and white positive correlations.  Correlations are calculated from 1953 using all of the QBO data (shown on the left) and using data just from the second half of the time series, 1979 (shown on the right).  The largest correlations for each figure are mentioned in the subheader of each figure.
}
     \label{f02}
\end{figure}

Both figures 1 and 2 also show correlations between the hurricane numbers and the QBO over the satellite era (1979-2006) where the intensity of the storms may be better measured.  We don't expect the difference in period to make much difference in the correlation plots if there is a consistent signal through the time series because the broad categorization of the storms as either category 1-5 or category 3-5 should not change dramatically due to small improvements in the intensity measurements. In general, though, the two periods do have differences.  The satellite era correlations emphasize a negative relationship with mid-stratospheric levels the year before.  This might imply that years with low hurricane activity are related to westerly QBO winds the winter before the hurricane season.  However, the 1953-2006 correlations seem to imply more positive correlations with the lower stratospheric QBO the winter before a hurricane season.  For this period, the late winter, March, provides the strongest correlations, indicating that a westerly QBO during March may imply a more active hurricane season for that same year.  These views are somewhat consistent because the October westerly QBO in the mid-stratosphere is anti-correlated with the March QBO in the lower-stratosphere at a value of -0.60.  But again, I emphasize that the correlations between the hurricane numbers and the QBO at any height or season are not significant and may be random rather than suggestive of any real relationship.\\
\\
However, because the QBO tends to be either westerly or easterly, without much time for a transition between states, perhaps we can more clearly see a relationship if we composite the hurricane numbers according to the phase of the QBO.  Figure 3 shows the results of compositing the hurricane numbers by phases of the QBO.\\
\\

\begin{figure}[!hb]
  \begin{center}
    \scalebox{0.5}{\includegraphics{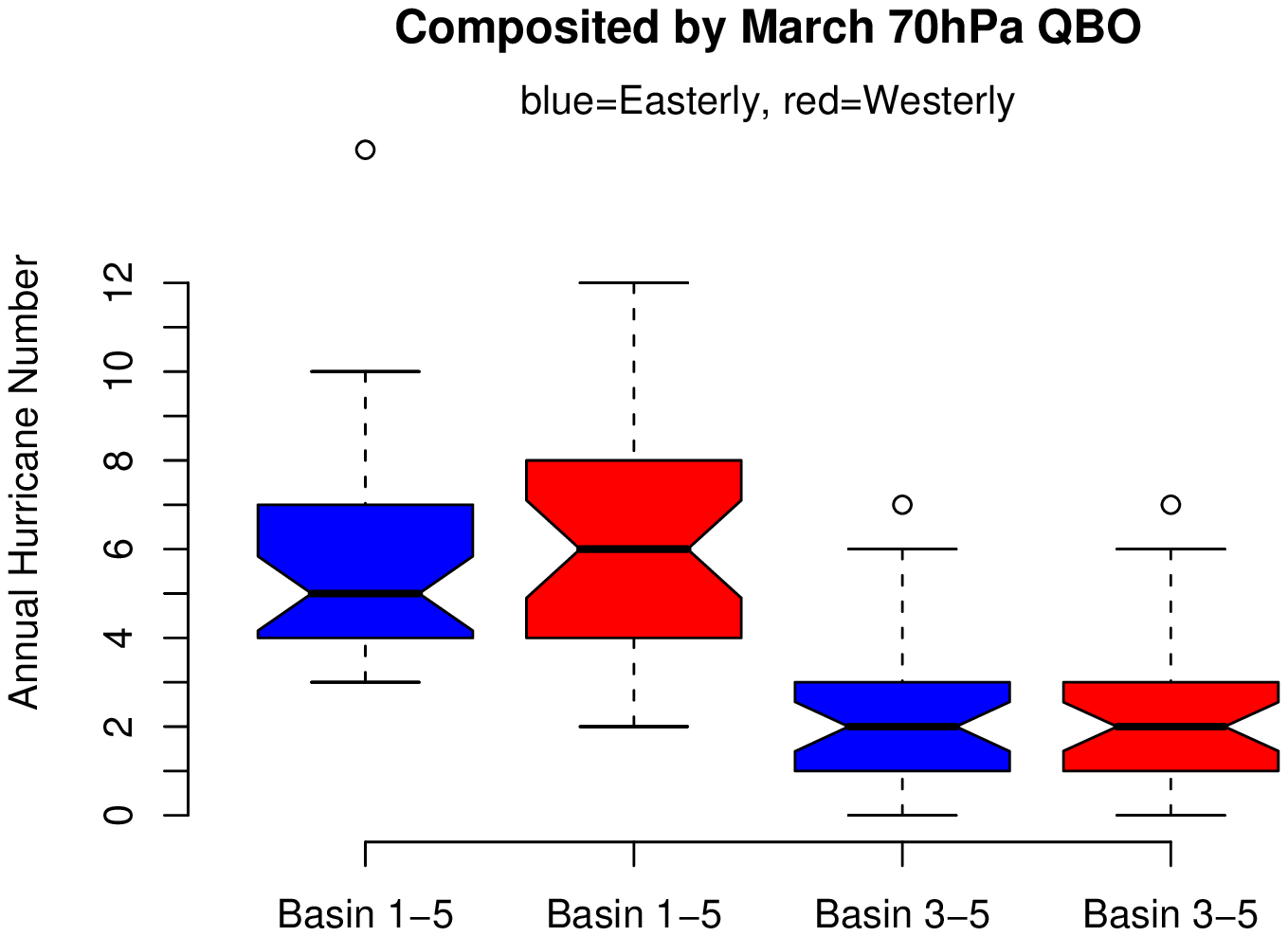}}
    \scalebox{0.5}{\includegraphics{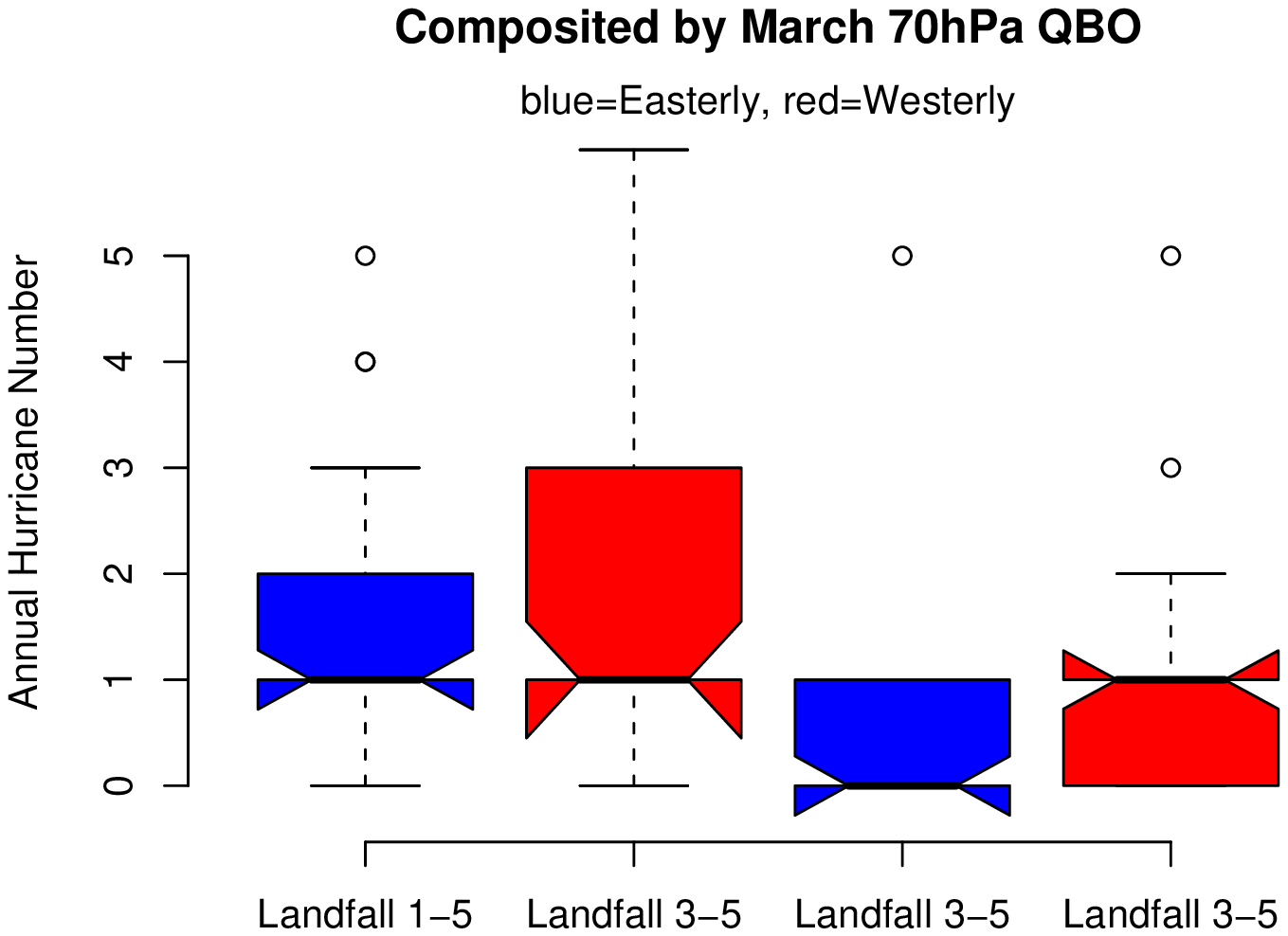}}
  \end{center}
    \caption{\textbf{Annual Hurricane Numbers composited according to the phase of the March 70hPa QBO one year ahead.}}
     \label{f03}
\end{figure}

\begin{figure}[!hb]
  \begin{center}
    \scalebox{0.5}{\includegraphics{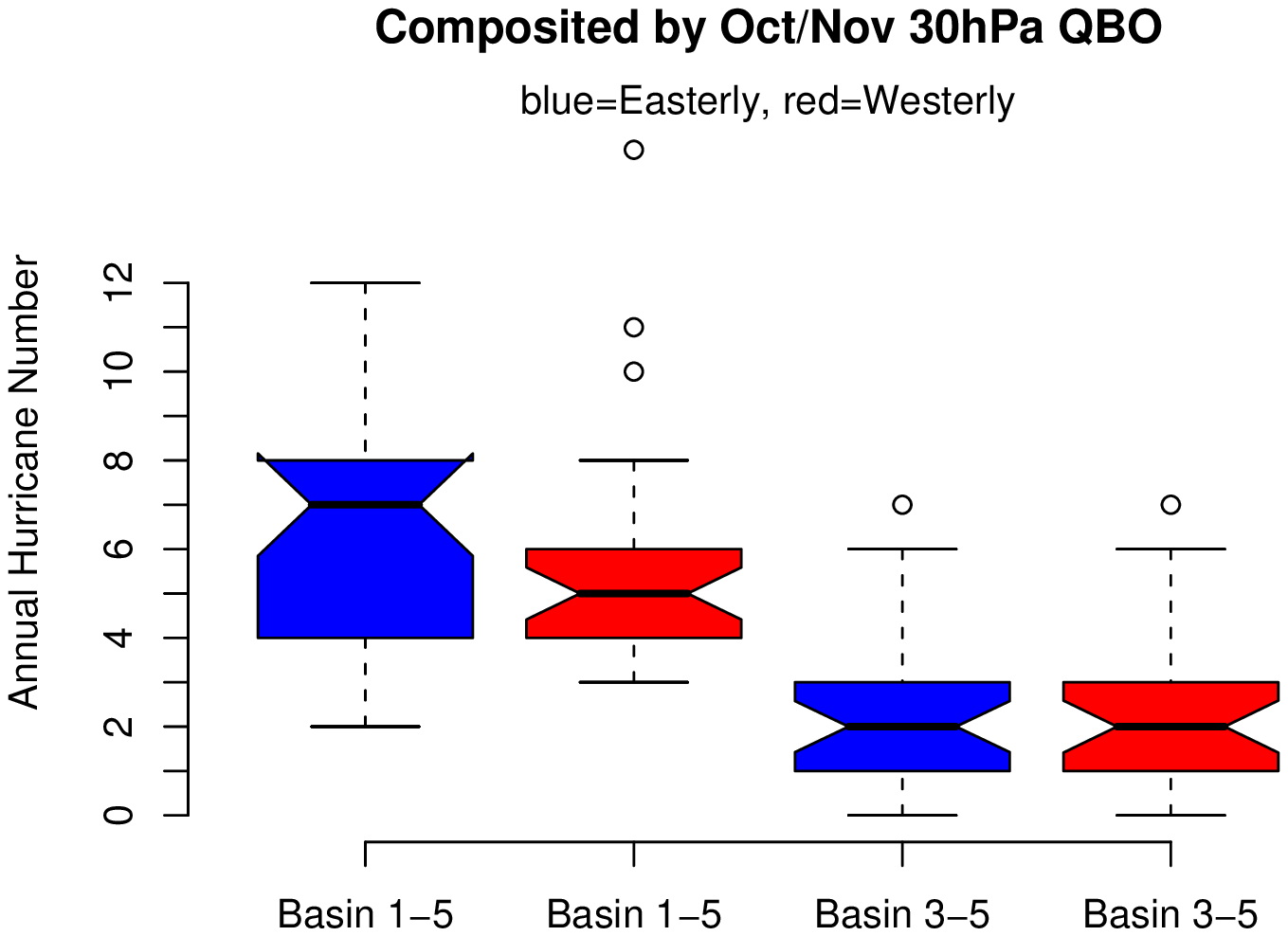}}
    \scalebox{0.5}{\includegraphics{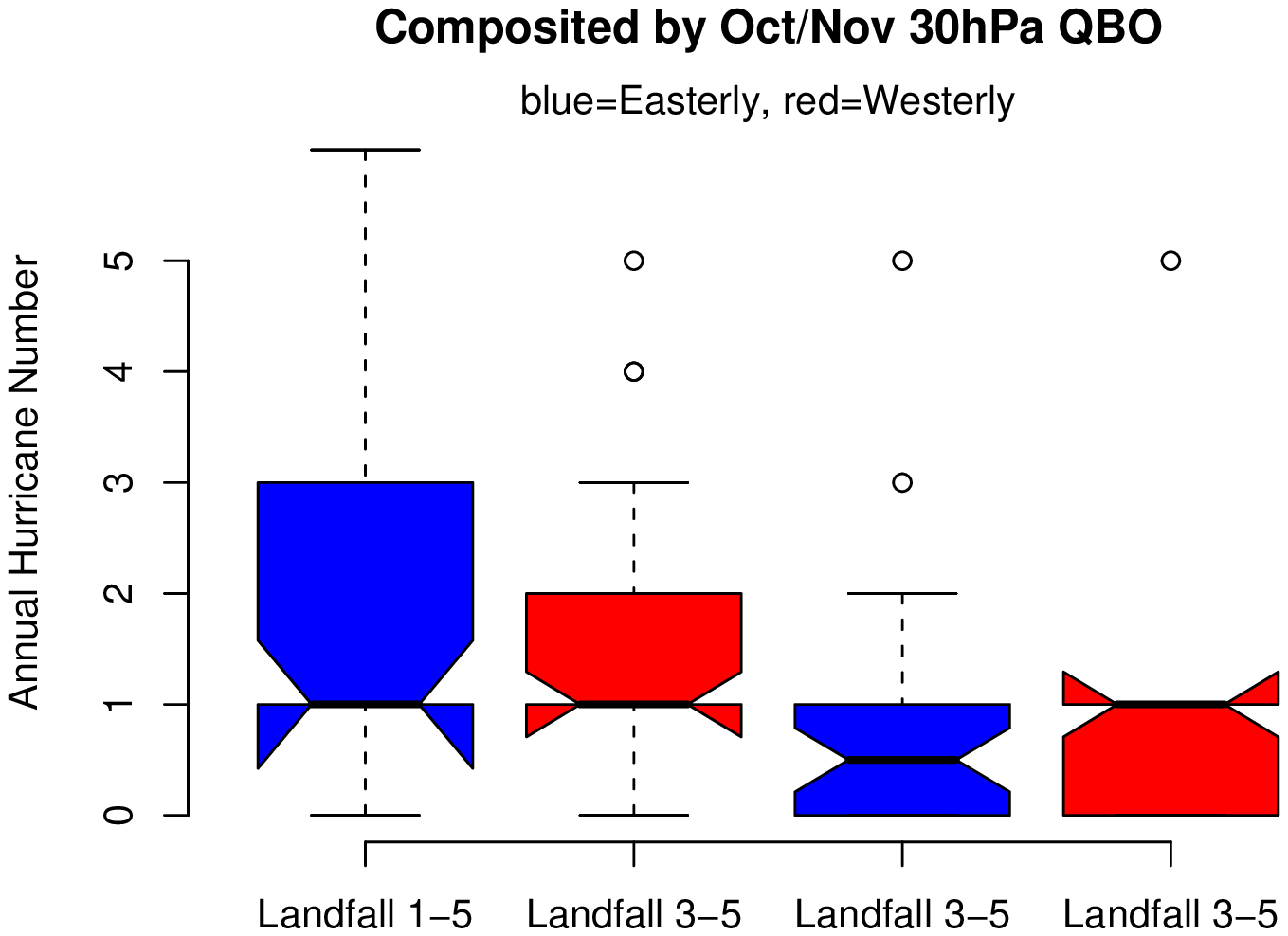}}
  \end{center}
    \caption{\textbf{Annual Hurricane Numbers composited according to the phase of the October-November averaged 30hPa QBO one year ahead.}}
     \label{f04}
\end{figure}

\begin{figure}[!hb]
  \begin{center}
    \scalebox{0.5}{\includegraphics{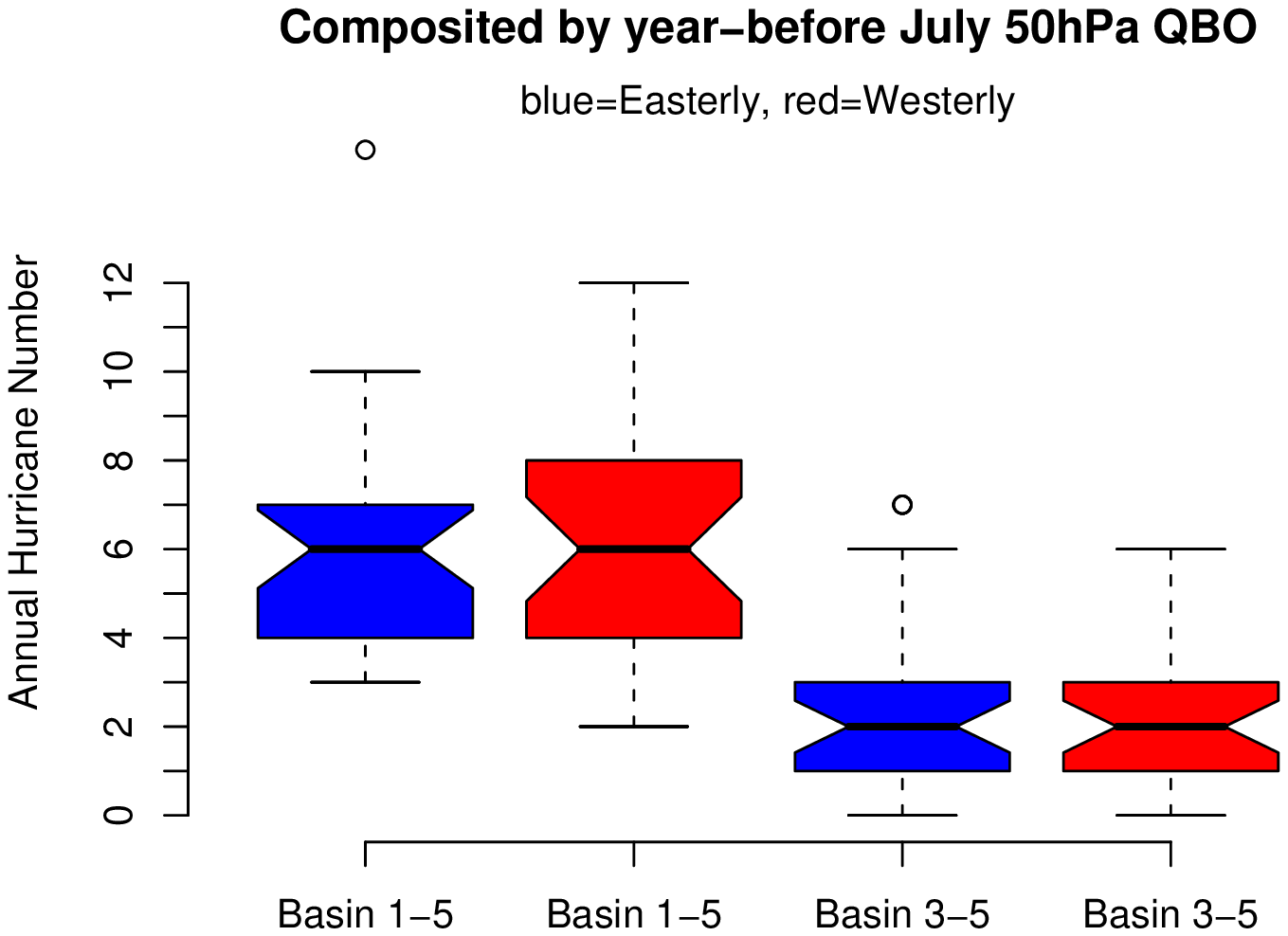}}
    \scalebox{0.5}{\includegraphics{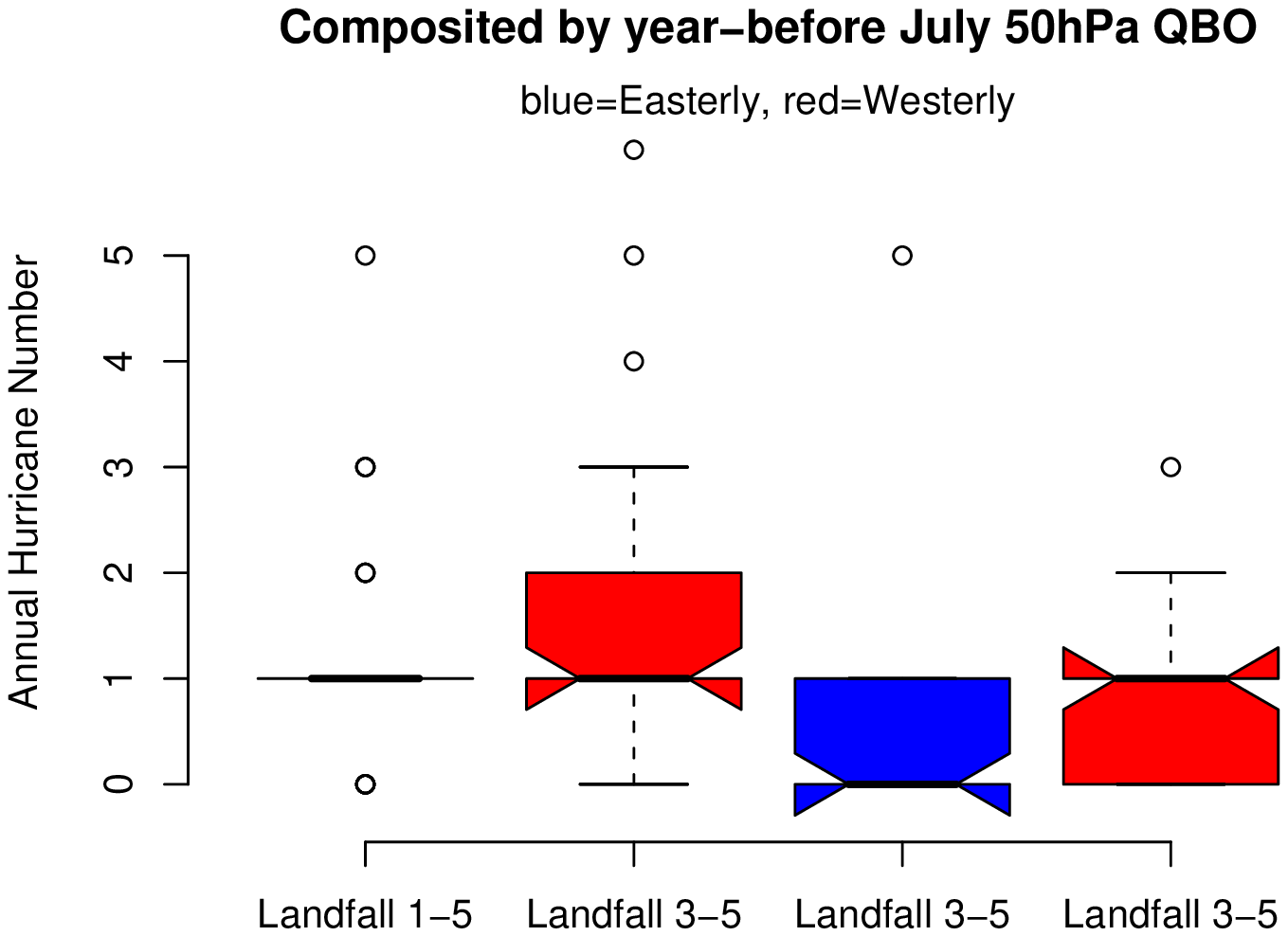}}
  \end{center}
    \caption{\textbf{Annual Hurricane Numbers composited according to the phase of the July 50hPa QBO one year ahead.}}
     \label{f05}
\end{figure}

Boxplots in figures 3, 4, 5 and 6 show the range of the hurricane numbers for various phases of the QBO.  The colored polygons (blue/red for the number of storms grouped by easterly/westerly QBO) indicate the 1st and 3rd quartile range and the solid horizontal line is the median.  The notches extend to +/-1.58 IQR/sqrt(n), where IQR is the inner quartile range and n is the number of years in each sample. This is based on the same calculations as the formula with 1.57 in Chambers et al. [1983, p. 62], given in McGill et al. [1978, p. 16]. There is an assumption of asymptotic normality of the median and roughly equal sample sizes for the two medians being compared, and the results are said to be rather insensitive to the underlying distributions of the samples. In this case, the sample sizes are roughly the same and Poisson simulations with sample sizes of 30 and lambda equal to the mean of either the basin numbers or the landfalling numbers show that the median of these distribution is close to normally distributed (using a Q-Q plot).  However, for the numbers of intense landfalling storms (cat3-5), there are not enough storms per year to give a normally distributed median.  For the other categories, this roughly gives a 95\% confidence interval for the difference in two medians.  If the notches of two plots do not overlap this is 'strong evidence' that the two medians differ [Chambers et al., 1983, p. 62].  We also note here that a comparison of the medians is much more robust to the influence of outliers than a comparison of means.\\
\\
In figure 3, we show the results for the hurricane numbers according to the phase of the March QBO at 70hPa.  This is the QBO index that had the largest correlation with the hurricane numbers when the entire 1953-2006 period is considered and from the correlation plots we expect that westerly winds at this level and time might precede a more active hurricane season.  The general result from figure 3 is that separating the hurricane seasons by this QBO index does not seem to define active and inactive seasons.  This is consistent with the fact that the correlations in figure 1 and 2 are not significantly different from zero.\\
  \\
  Figure 4 shows the composite according to the phase of the October-November averaged QBO at 30hPa (the index that had the highest correlation for the satellite era) and figure 5 shows the composite according to the phase of the year-before July QBO at 50hPa (which is the index that Klotzbach and Gray used for their seasonal forecasts [Klotzbach and Gray, 2007]).\\
\\
 A comparison of the easterly and westerly groups indicates that the basin 1-5 storms and the landfalling 3-5 storms do seem to differ according to the phase of the QBO one year ahead.  In particular, the landfalling cat 3-5 numbers indicate a difference between easterly and westerly years despite no difference in landfalling cat 1-5 storms or in basin cat 3-5 storms.  Most likely this is a sampling issue as opposed to any real difference.  The difference in the basin cat 1-5 storms can be seen most clearly in the October-November average of the 30hPa QBO.  When these winds are easterly, the median number of hurricanes in the Atlantic basin tends to be higher in the following year compared to when these winds are westerly.  It should be noted that there is considerable overlap between the two groups and it is only a difference in where the median lies that is significant.\\
 \\
 Furthermore, when we then ask if this gives us any hope in distinguishing the annual average of the following five years according to the phase of the QBO, we find that it does not.  Figure 6 shows the annual average numbers of hurricanes in the 5-years following an easterly or westerly October-November averaged 30hPa QBO.  From this we see that any possible difference in hurricane numbers due to the phase of the QBO is lost, or averaged out, when we try to predict on longer time scales.  None of the composite groups are different.\\

\begin{figure}[!hb]
  \begin{center}
    \scalebox{1.}{\includegraphics{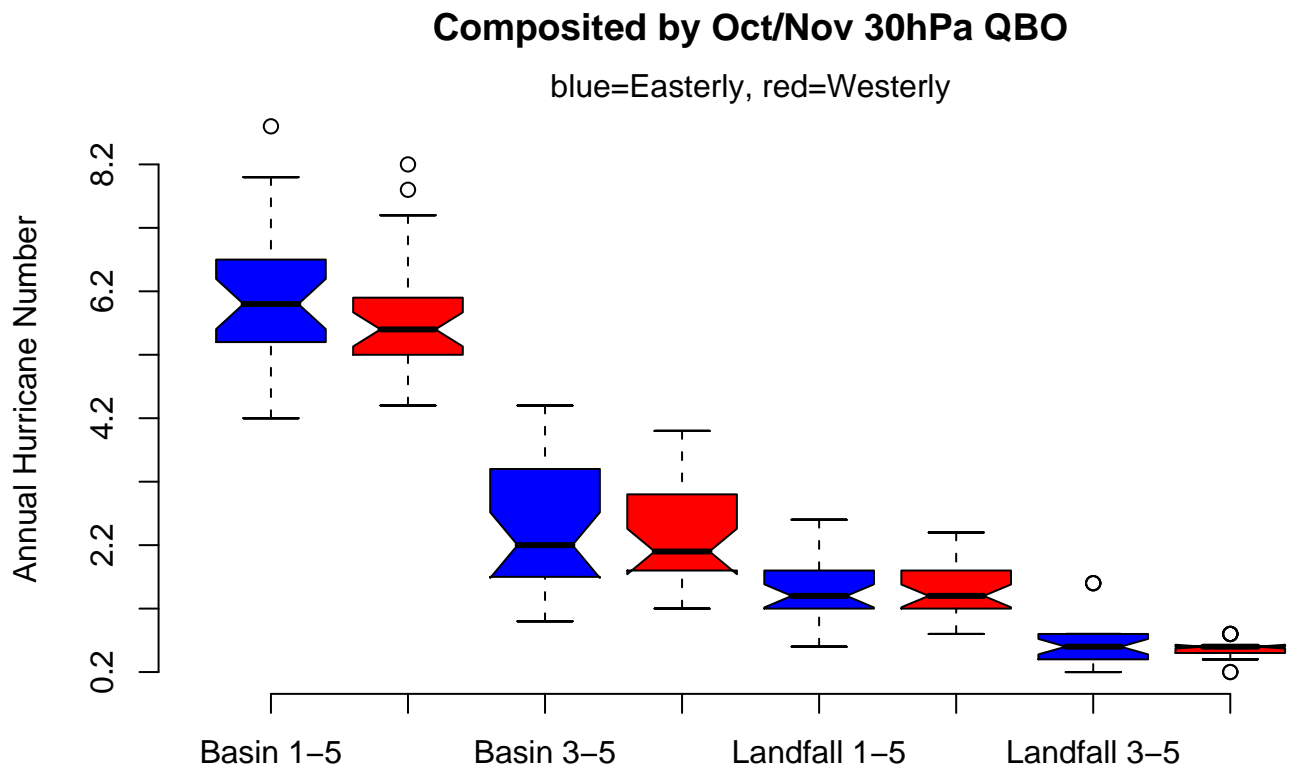}}
  \end{center}
    \caption{5-year mean Annual Hurricane Numbers composited according to the phase of the October-November averaged 30 hPa QBO one year before the 5-year mean.}
     \label{f06}
\end{figure}

Up to now, we have shown that the medians of the two groups are not statistically different. Since this is a more robust measure than differences in means (which can be unduly affected by outliers) we do not expect that the difference in the mean easterly/westerly composites would be useful in predicting hurricane numbers.  However, the difference in means is sometimes still used as a predictor [Shapiro, 1989 and Klotzbach and Gray, 2007].  So here we confirm our hypothesis by showing predictions using the phase of the QBO and comparing the errors of these predictions to the error in using the out-of-sample mean as a prediction.  Figure 8 shows the out-of-sample hindcast predictions using the phase of the year-before July 50 hPa QBO (as used in Klotzbach and Gray [2007]) alongside the actual hurricane numbers and an out-of-sample mean.  Table 1 shows the root mean squared errors (RMSEs) of these forecasts and from this we can see that there is a higher error associated with using the QBO as a predictor than when using the climatological mean.\\

\begin{figure}[!hb]
  \begin{center}
    \scalebox{0.5}{\includegraphics{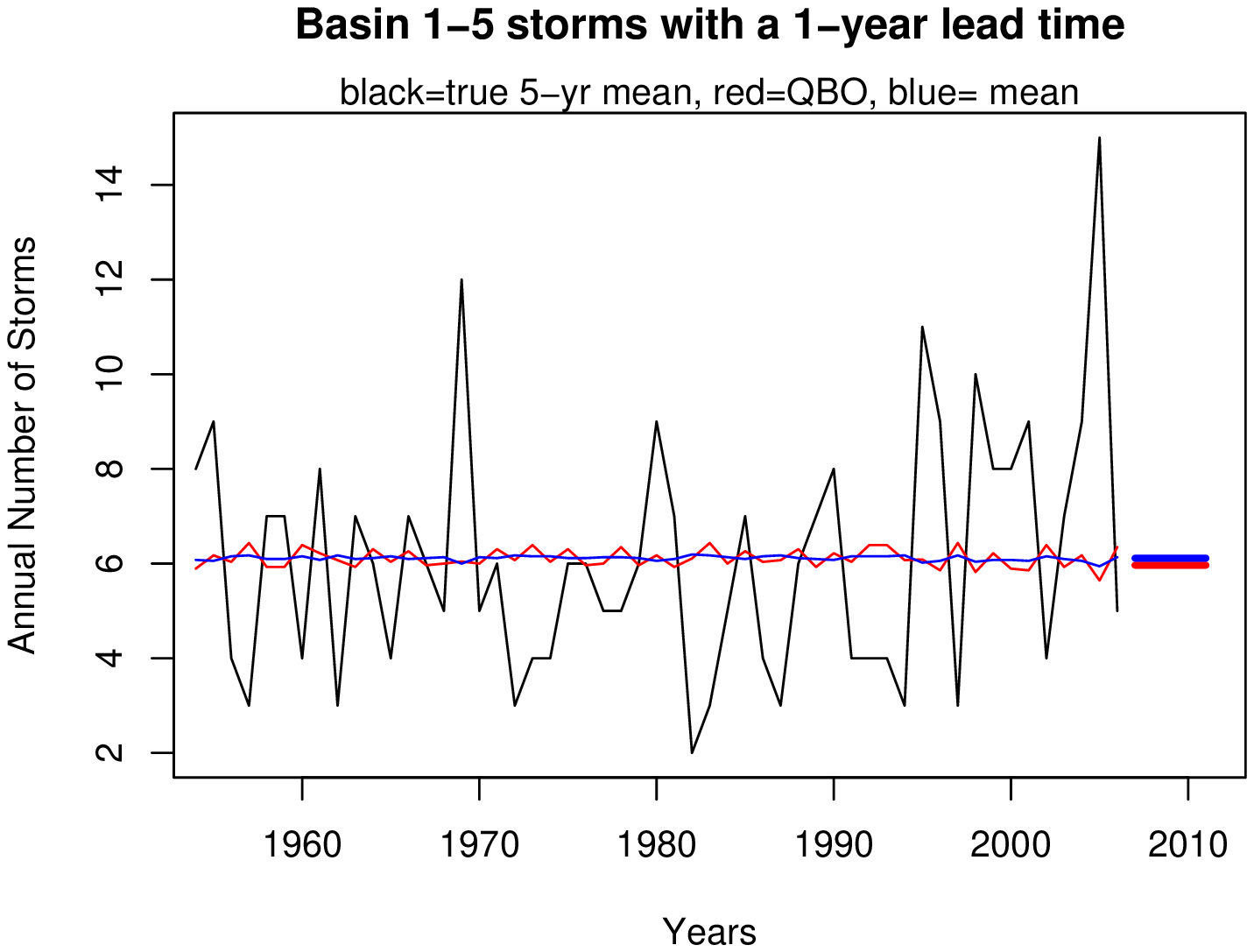}}
    \scalebox{0.5}{\includegraphics{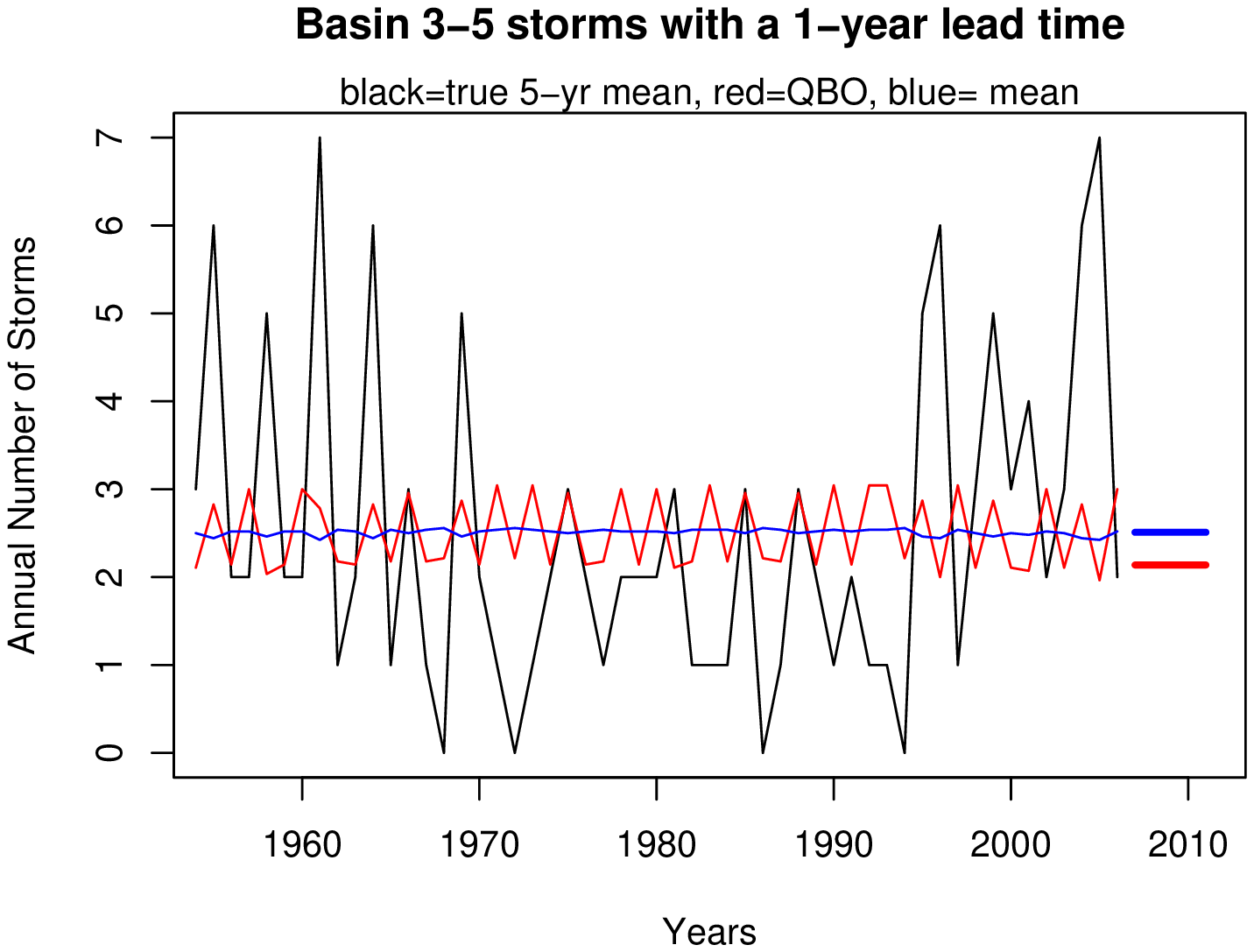}}
  \end{center}
    \caption{1-year ahead predictions of yearly basin hurricane numbers (cat1-5 top and cat 3-5 bottom).  The red line shows the out-of-sample hindcasts using the July 50hPa QBO and a blue line shows the out of sample mean prediction.  The subsequent predictions for 2007 are shown as bold lines at the end of the time series.}
     \label{f07}
\end{figure}

\begin{figure}[!hb]
  \begin{center}
    \scalebox{0.5}{\includegraphics{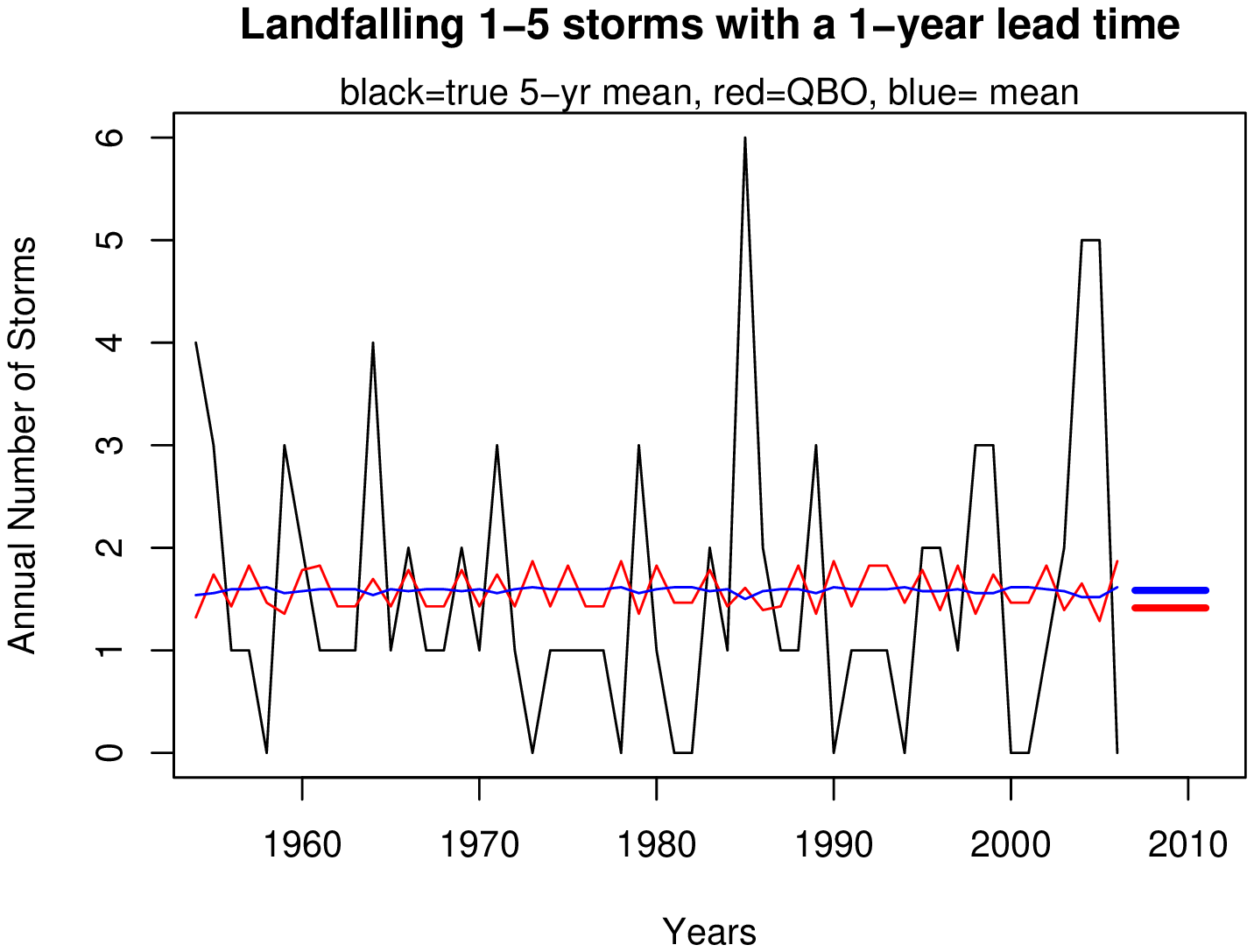}}
    \scalebox{0.5}{\includegraphics{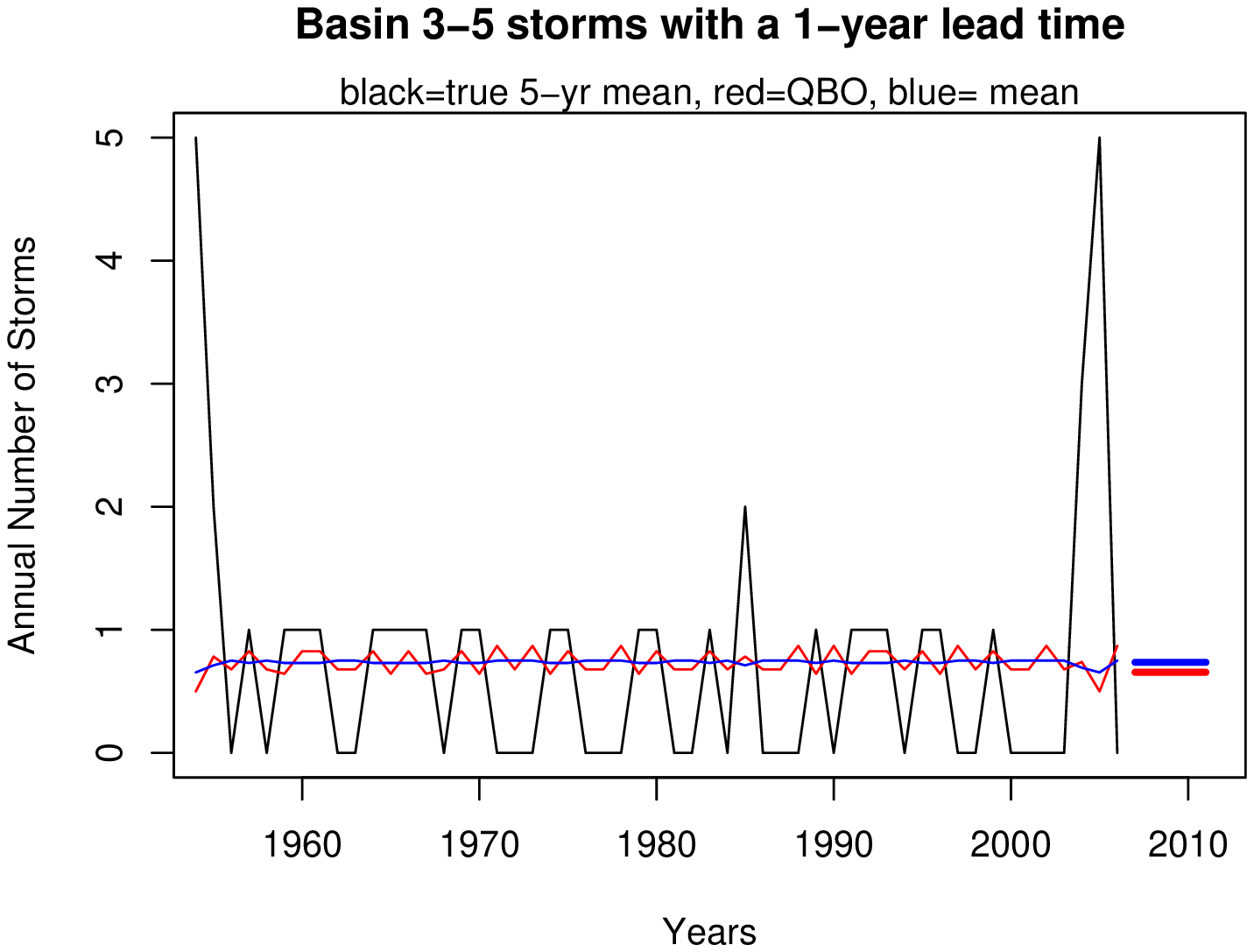}}
  \end{center}
    \caption{1-year ahead predictions of yearly landfalling hurricane numbers (cat1-5 top and cat 3-5 bottom).  The red line shows the out-of-sample hindcasts using the July 50hPa QBO and a blue line shows the out of sample mean prediction.  The subsequent predictions for 2007 are shown as bold lines at the end of the time series.}
     \label{f08}
\end{figure}

\begin{table}[!h]\centering\begin{tabular}{|c|c|c|}
  \hline
  Prediction Method & Landfalling RMSE & Basin RMSE \\

\hline
QBO for cat 1-5   &   2.061 & 7.209 \\
\hline
mean for cat 1-5 & 2.016 & 6.964 \\
\hline
QBO for cat 3-5     & 1.216 & 3.476 \\
\hline
mean for cat 3-5 & 1.182 & 3.513 \\
\hline

\end{tabular}\caption{RMSE of the hindcasts of annual mean hurricane numbers for a 1-year lead time. The QBO index here refers to the year-before July 50hPa winds.}\label{t01}
\end{table}

From Table one we see that, although neither of the prediction methods account for the longer term variability, using a long term mean actually does a better job of predicting hurricane numbers than using the July QBO index.  The one exception seems to surprisingly be the basin, category 3-5 storms.  However, as expected, we lose any of this apparent skill when we try to predict an annual average for 5-years ahead.  Root mean squared errors for the Landfalling and Basin predictions are shown in table 2.  The QBO index used in this case is the October-November averaged 30hPa winds which gives slightly better results than the other indices but still does not perform as well as a long term mean.  This table also contains errors from what is called "history only" predictions.  In these cases, the hindcasting experiments are performed using only information that is older than the period to be hindcast. One interesting thing about these predictions is that the QBO predictions tend to converge towards the mean prediction over time as can be seen in figures 9 and 10 and shown explicitly in figure 11.  \\

\begin{table}[!h]\centering\begin{tabular}{|c|c|c|c|c|}
  \hline
  Prediction Method & LF RMSE & Basin RMSE & History Only LF RMSE & History Only Basin RMSE\\

\hline
QBO for cat 1-5   &   0.250 & 1.213 & 0.264 &  1.275\\
\hline
mean for cat 1-5 & 0.243 & 1.182 & 0.256 & 1.229\\
\hline
QBO for cat 3-5     & 0.100 & 0.854 & 0.099 & 0.917\\
\hline
mean for cat 3-5 & 0.0987 & 0.829 & 0.0926 & 0.871\\
\hline

\end{tabular}\caption{RMSE of the hindcasts of annual mean hurricane numbers over a 5-year interval.  The QBO index here refers to the October-November 30hPa winds.}\label{t02}
\end{table}

\begin{figure}[!hb]
  \begin{center}
    \scalebox{0.5}{\includegraphics{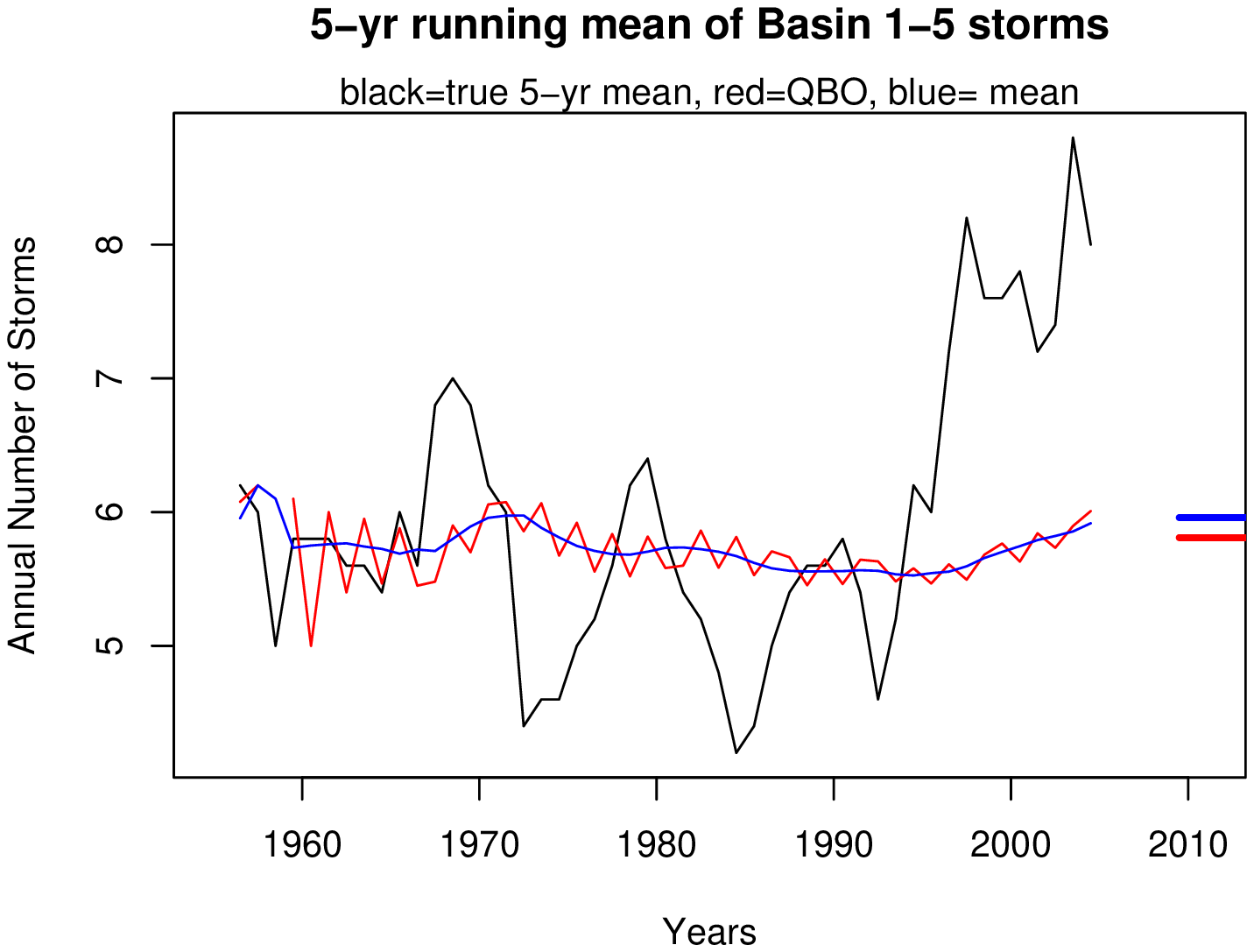}}
    \scalebox{0.5}{\includegraphics{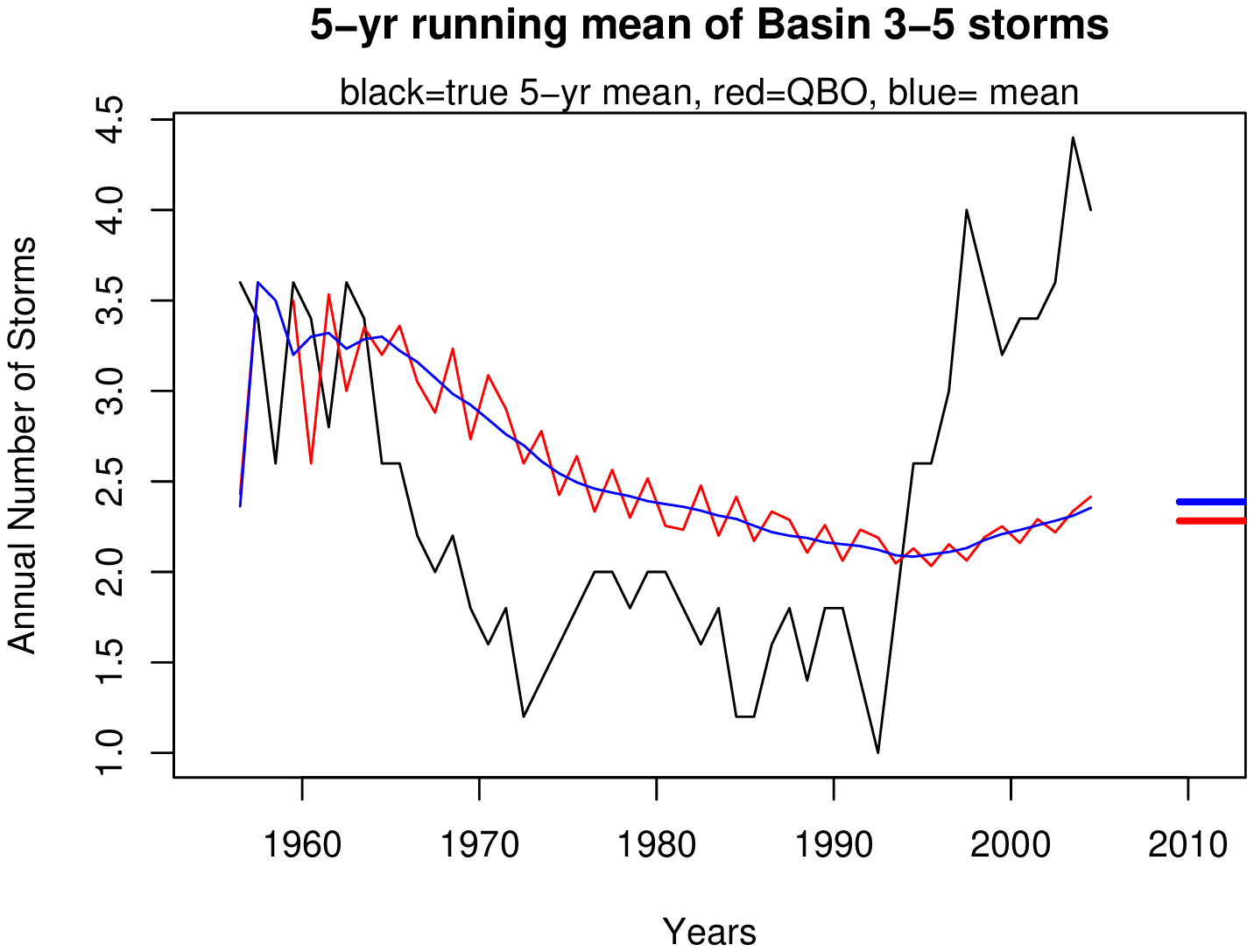}}
  \end{center}
    \caption{}
     \label{f09}
\end{figure}

\begin{figure}[!hb]
  \begin{center}
    \scalebox{0.5}{\includegraphics{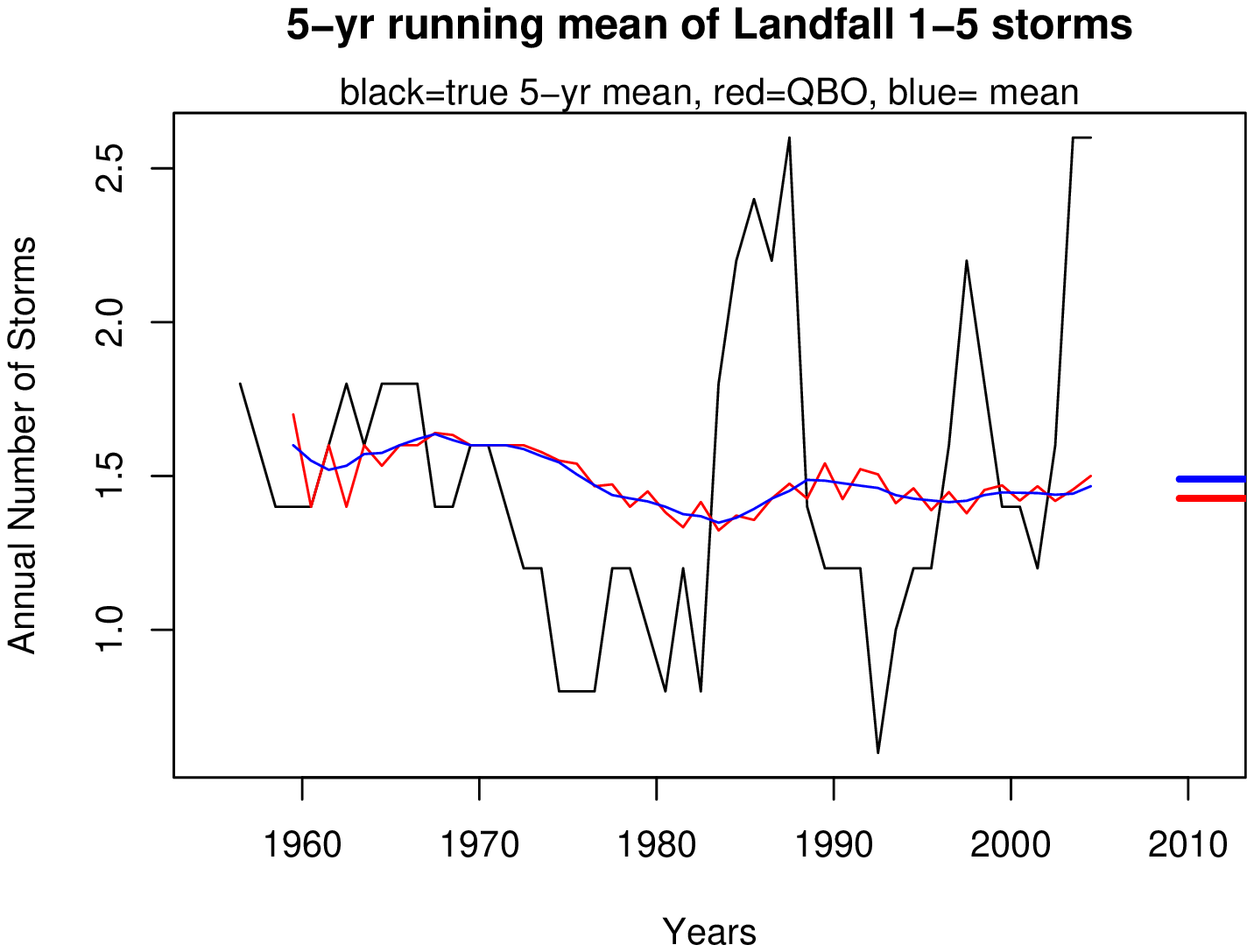}}
    \scalebox{0.5}{\includegraphics{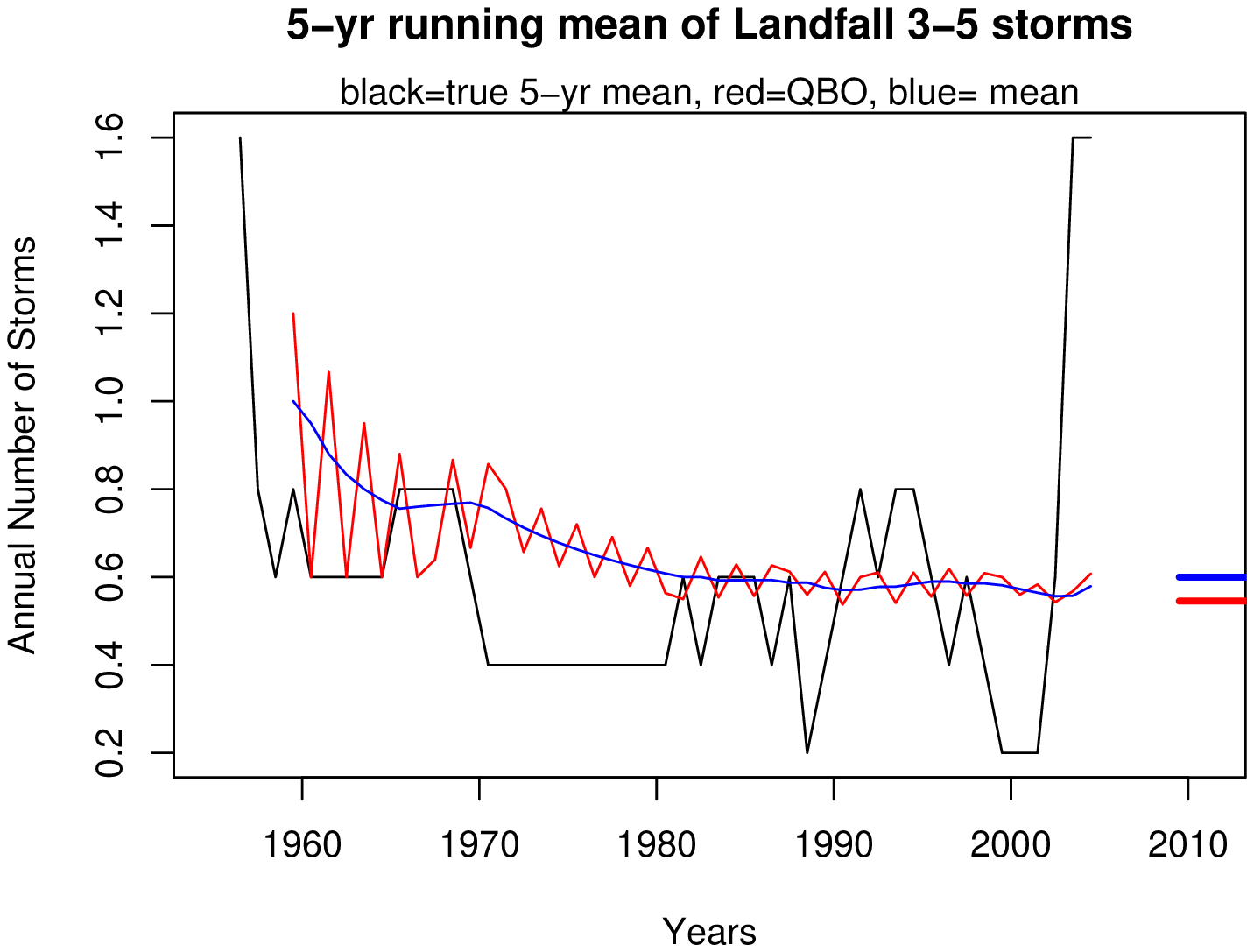}}
  \end{center}
    \caption{5-year annual averaged landfalling numbers (cat1-5 top and cat 3-5 bottom).  The red line shows the out-of-sample hindcasts using the Oct-Nov 30hPa QBO and a blue line shows the out of sample mean prediction.  Values for the 5-year mean observations and predictions are all plotted at the center date of the 5-year mean.}
     \label{f10}
\end{figure}

\begin{figure}[!hb]
  \begin{center}
    \scalebox{0.5}{\includegraphics{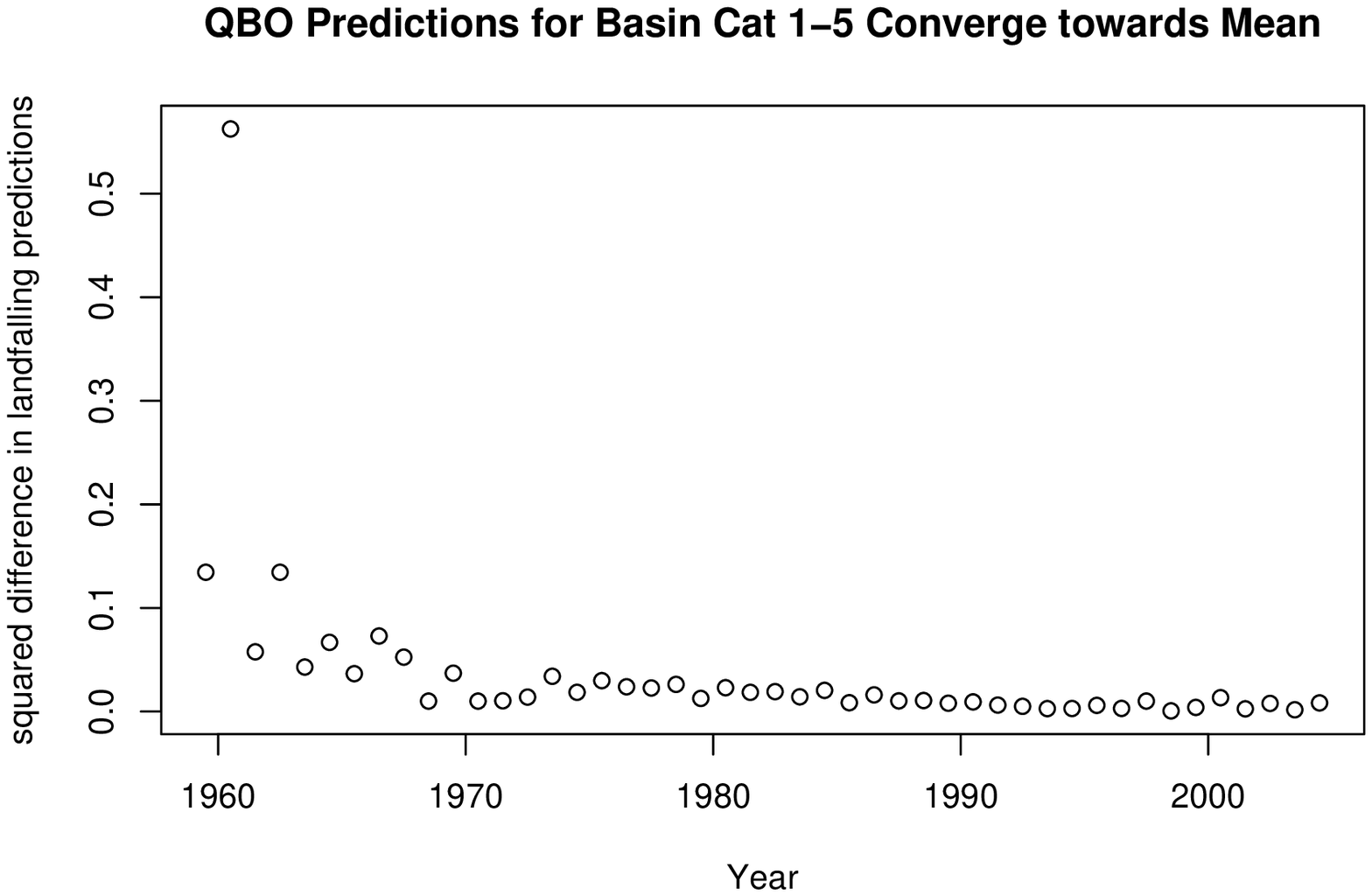}}
    \scalebox{0.5}{\includegraphics{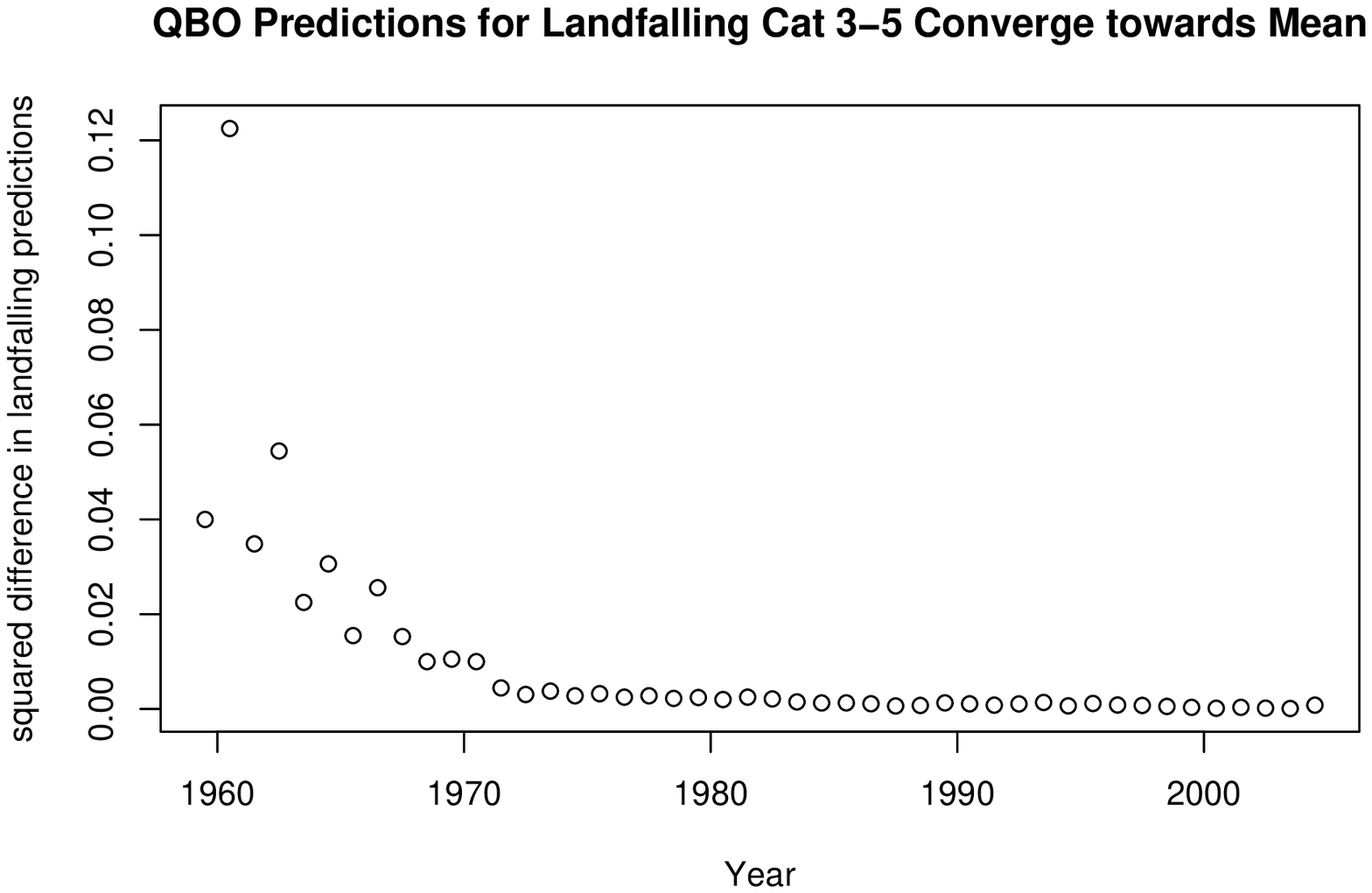}}
  \end{center}
    \caption{Squared difference between the QBO predictions and the mean.}
     \label{f11}
\end{figure}

\section{Discussion}

In this simple study we see no difference in the five year mean number of hurricanes when they are categorized by the phases of the QBO.  Furthermore predictions of hurricane numbers using the phase of the QBO as a predictor have larger errors than when just a climatological mean is used for prediction.

\section{bibliography}
Chambers, J. M., Cleveland, W. S., Kleiner, B. and Tukey, P. A. (1983) Graphical Methods for Data Analysis. Wadsworth \& Brooks/Cole.\\
\\
Gray, W.M., C.W. Landsea, P.W. Mielke and K.J. Berry (1992), Predicting Atlantic seasonal hurricane activity 6-11 months in advance. Wea. Forecasting, 7, 440-455.\\
\\
Jarvinen, B.R., C.J. Neumann and M.A.S. Davis (1984), A tropical cyclone data tape for the North Atlantic Basin, 1886-1983: contents, limitations and uses; NOAA Technical Memorandum NWSNHC, 22, p.21.\\
\\
Klotzbach, P.J. and W.M. Gray (2007), Extended Range Forecast of Atlantic Seasonal Hurricane Activity and U.S. Landfall Strike Probability for 2007, University of Colorado.\\
\\
Hollander, M. and D.A. Wolfe (1973), Nonparametric statistical inference. New York: John Wiley \& Sons. Pages 185–194.\\
\\
McGill, R., J.W. Tukey and W.A. Larsen (1978): Variations of box plots. The American Statistician 32, 12–16.\\
\\
Naujokat, B., (1986): An update of the observed quasi-biennial oscillation of the stratospheric winds over the tropics. J. Atmos. Sci., 43, 1873-1877.\\
\\
Neumann, C.J. (1993): "Global Overview - Chapter 1":  Global Guide to Tropical Cyclone Forecasting, WMO/TC-No. 560, Report No. TCP-31, World Meteorological Organization; Geneva, Switzerland.\\
\\
Shapiro, L. J., 1989: The relationship of the quasi-biennial oscillation to Atlantic tropical storm activity. Mon. Wea. Rev., 117, 1545-1552.\\
\\

\end{document}